# Spin configurations in hard–soft coupled bilayer systems: from rigid magnet to exchange spring transitions

N. Sousa, A. Apolinario, P. M. S. Monteiro, D. S. Schmool
*IFIMUP-IN and Departement of Physics, University of Porto,
Rua Campo Alegre 687, I - 43124 Porto, Portugal*

F. Vernay, H. Kachkachi
*LAMPS, Université de Perpignan Via Domitia, 52 Avenue Paul Alduy, F-66860 Perpignan, France*

F. Casoli, F. Albertini
*Istituto IMEM-CNR, Parco Area delle Scienze 37/A, 43010 Parma, Italy*

We investigate equilibrium properties of an exchange-spring magnetic system constituted of a soft layer (*e.g.* Fe) of a given thickness on top of a hard magnetic layer (*e.g.* FePt). The magnetization profile $M(z)$ as a function of the atomic position ranging from the bottom of the hard layer to the top of the soft layer is obtained in two cases with regard to the hard layer: i) in the case of a rigid interface (the FePt layer is a single layer), the profile is obtained analytically as the exact solution of a sine-Gordon equation with Cauchy's boundary conditions. Additional numerical simulations also confirm this result. Asymptotic expressions of $M(z)$ show a linear behavior near the bottom and the top of the soft layer. In addition, a critical value of the number of atomic planes in the soft layer, that is necessary for the onset of spin deviations, is obtained in terms of the anisotropy and exchange coupling between the adjacent plane in the soft layer. ii) in the case of a relaxed interface (the FePt layer is a multilayer), the magnetization profile is obtained numerically for various Fe and FePt films thicknesses and applied field.

## I. INTRODUCTION

The proposal of hard–soft coupled or exchange spring (ES) systems as permanent magnets with high performances dates back to 1991, but after 2005 a renewed interest in these systems has increased due to their potential use as magnetic recording media with high thermal stability and reduced switching field[1–3]. While the first theoretical models considered the overall magnetic behavior of the system and calculated the maximum energy product values[4–6], recent theoretical works have focused on the performances of the systems in magnetic recording, *e.g.*, thermal stability, switching field, and switching time[7–9].

The ES system shows new magnetic properties with respect to its hard and soft constituent components, being characterized by the competition of the components anisotropy contributions (both magneto-crystalline and shape) and strong exchange coupling between the two phases, which produces a nonuniform spin orientation predominantly in the soft layer.

Bilayers with the in-plane easy direction have been considered as model systems in many works[4–6], however, due to the interest in magnetic recording applications, recent interest has been directed towards ES systems with perpendicular magnetic anisotropy or exchange-coupled composite media made of separated grains[7–9]. For this reason we have recently studied Fe/FePt bilayers, where the hard FePt layer has perpendicular orientation[10,11]. Different magnetic regimes can be identified upon varying the Fe layer thickness. The limit of the rigid magnet (RM) regime (*i.e.*, where the soft layer reversal is collinear with the hard layer) is a function of the sample interface morphology and hard/soft coupling intensity[10,11]. With increasing Fe thickness, the bilayers pass from RM to ES behavior, with a reversible portion of the demagnetization curve.

An important issue which is common to all exchange-coupled systems concerns the description of the spin behavior in the coupled layers. In the present work we have developed a one-dimensional spin model ($1D$ model) analysis of ES bilayer system, which takes the case of a hard phase with perpendicular anisotropy. To study the static case of the spin configuration, we start by establishing the various energy contributions based on the magnetic energy of the coupled layers. Applying the equilibrium condition, we then minimize the energy on a spin by spin basis in order to evaluate the equilibrium orientation as a function of position. We have performed calculations using two different interface conditions:

1. Rigid interface; where the first spin of the Fe (soft) layer is held rigidly in the direction of the FePt easy axis, perpendicular to the film plane.

2. Relaxed interface; where the spin configuration allows a rotation of FePt spins such that the domain wall (DW) produced in the ES system can penetrate both magnetic layers to varying degrees, depending on their individual magnetic properties. This is not the case for the rigid interface condition, where the spin rotation or DW is located only in the Fe (soft) layer.

In order to verify the predictions of the model we have performed both analytical and numerical studies of the angle variation of the individual spins as a function of the number of spins. An issue in this numerical simulation concerns the demonstration, made in the present work, of the validity of the one-dimensional assumption for the perpendicular situation. We have considered

- the variation of the number of spins (in both magnetic layers) to verify the transition between the two regimes (RM - ES).
- the variation of the relative strength of the Fe anisotropy constant (in case of the rigid interface approach) and FePt anisotropy in the case of the relaxed interface approach.
- various strengths and directions of a static external magnetic field.

To further test the model, we have made a comparison to analytical results and corresponding simulations using the OOMMF software[12], routinely used in micromagnetic simulations.

Our recent ferromagnetic resonance (FMR) measurements on FePt/Fe layers showed a strong uniaxial anisotropy induced in the Fe layer via the exchange coupling with the FePt film[13]. Exploiting this result, we have developed a model as close as possible to the real system. The calculations we present here are in fact based on the experimental values measured for the structural and magnetic parameters, that is, lattice spacing, saturation magnetization and anisotropy constant in the epitaxial FePt(001)/Fe(110) system[10]. This aspect, together with the ability to obtain the magnetic phase diagram and spin configuration as a function of a varying field applied at different angles, give an added value to this work. They in fact allow a wide predictive capability on the results of magnetic measurements on real systems[10,13,14] and magnetic media performances in the presence of the head field[9].

The paper is organized as follows: in section II we describe the physical system and introduce the model under consideration; then to investigate the underlying physics we consider in section III the limiting case of a rigid interface for which we provide an exact analytical solution and check it against numerical simulations. Section IV is devoted to the numerical treatment of the relaxed interface case under a magnetic field. In section V we discuss the experimental relevance of the present work and future ones. Section VI contains our concluding remarks.

## II. STATEMENT OF THE PROBLEM AND MODEL

The experimental situation is as follows: we consider a multilayer system consisting of two different epitaxial films; iron layers which take a (110) orientation stacked on ferromagnetic FePt layers with (100) orientation[10] with respect to the sample plane as shown in Fig. 1. From a structural point of view, the iron film has a BCC structure and a given spin (at the $n^{th}$ layer) interacts with 4 neighboring spins within its atomic layer (red spheres in Fig.1) and with 4 others out of plane (gray spheres), i.e., two in the $n+1$ plane and two in the $n-1$ plane. In the BCC iron structure, the exchange parameter is $J_{\text{Fe}} = 1.44 \times 10^{-21} J$ and the lattice parameter is $a = 2.86 \text{Å}$ [15]. The iron platinum film has a face-centered tetragonal (FCT) $L1_0$, structure with an exchange constant of $J_{\text{FePt}} = 2.07 \times 10^{-21} J$, and with lattice parameters $a = 3.86 \text{Å}$, $c = 3.71 \text{Å}$ [10,16]. In order to assess the respective magnetic anisotropies of the layers, we have initially considered the Fe and FePt anisotropy values to be variable in magnitude. The anisotropy in the FePt layer is considered to be larger than that of the Fe layer ($D_{\text{FePt}} > D_{\text{Fe}}$), in the following we will refer to them as, respectively, $D_{\text{H}}$ and $D_{\text{S}}$, the subscript H standing for hard and S for soft. Similarly, the Fe film will be henceforth referred to as the soft layer (SL) and the FePt film as the hard layer (HL).

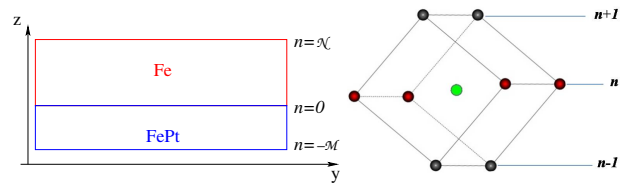

Figure 1: a) Representation of the system with N layers. b) BCC iron structure, where the red spheres represents the spins in the same plane, and the gray spheres represents spins in different atomic planes.

Since the FePt has an FCT structure for which we consider the (100) orientation, each spin in a given atomic plane is exchange coupled to only 2 neighbors in the upper plane and 2 in the lower plane.

Our aim here is to provide a theoretical formulation of the ES problem in the simplest way and set up the problem for the study of its dynamical behavior and, in particular, to investigate its excitations modes and their behavior upon varying its intrinsic physical characteristics and external fields. More precisely, we shall determine the magnetization profile through the whole thickness of the system, starting from the fixed layer at the bottom end of the HL up to the loose spins at the top of the SL. To do so the multilayer system is mapped onto a discrete problem: a stack of $\mathcal{N} + \mathcal{M}$ magnetic atomic layers each represented by a (normalized) magnetic moment $\mathbf{S}_n$ with $n = -\mathcal{M}, ..., 0, ..., \mathcal{N}$ and $\|\mathbf{S}_n\| = 1$ (see Fig. 1). Layers from $-\mathcal{M}$ to 0 correspond to the HL, while those ranging from $n = 1$ to $n = \mathcal{N}$ belong to the SL. The atomic layer $n = 0$ corresponds to the interface HL/SL. For the HL, the anisotropy axis is taken along the $z$ direction, normal to the layer plane; the corresponding anisotropy constant is $D_{\text{H}}$. On the other hand, each layer labeled by $n = 1, ..., \mathcal{N}$ has a uniaxial (shape) anisotropy with easy axis lying in the $xy$-plane. To be specific, we take it here to be along the $y$ direction; the anisotropy constant is $D_S$. In addition, all layers are subject to an external DC-field applied in the $yz$-plane.

We further assume that the lateral dimensions of the atomic planes are large enough so as to neglect boundary effects and to assume that atomic spins lying in the same atomic plane behave coherently. More precisely, we assume that under the effect of the intra-plane ferromagnetic exchange coupling and anisotropy, each plane behaves as a massive ferromagnet. This means that within an atomic layer all spins are parallel to each other and make the same polar angle $\theta_n$ with the $z$-axis. As such the angle deviation is assumed to vary only between adjacent planes and not within the planes.

Consequently, our model Hamiltonian reads

$$\begin{aligned}\mathcal{H} = &- (g\mu_B) \mathbf{H} \cdot \sum_{n=-\mathcal{M}}^{\mathcal{N}} \mathbf{S}_n \\ &- D_H \sum_{n=-\mathcal{M}}^{0} (S_n^z)^2 - D_S \sum_{n=1}^{\mathcal{N}} (S_n^y)^2 \\ &- \frac{1}{2} \sum_{n=-\mathcal{M}+1}^{\mathcal{N}-1} J_n \mathbf{S}_n \cdot (\mathbf{S}_{n+1} + \mathbf{S}_{n-1}) - \frac{J_{Fe}}{2} \mathbf{S}_{\mathcal{N}-1} \cdot \mathbf{S}_{\mathcal{N}} \\ &- \frac{J_{FePt}}{2} \mathbf{S}_{-\mathcal{M}+1} \cdot \mathbf{S}_{-\mathcal{M}} - \frac{J_0}{2} \mathbf{S}_0 \cdot \mathbf{S}_1.\end{aligned} \quad (1)$$

where the first term represents the Zeeman energy, the second and third are anisotropy contributions, and the last two lines are the total exchange energy comprising the coupling between all adjacent planes, including the bottom and top layers and the interface between the HL and SL. $J_0$ is the coupling between the last atomic plane in the HL and the first in the SL, $i.e.$ the interface between HL and SL.

In the following we shall consider two situations regarding the HL, either as a rigid magnet represented by a single macroscopic magnetic moment or as a stack of atomic layers. These two cases correspond to the respective situations of a rigid and relaxed interface.

### III. RIGID INTERFACE

In the present case the hard magnetic film is considered as a single layer represented by a macroscopic magnetic moment $\mathbf{S}_0$ pinned along its anisotropy axis so that $n = 0, 1, \ldots, \mathcal{N}$ [see Eq. (1)]. The magnetic moment of this HL is supposed to be infinitely rigid and thus unaffected by an applied magnetic field, leading to $\theta_0 = 0$. In addition, for the sake of simplicity, the problem is further simplified by assuming uniform exchange coupling, $i.e.$, $J_n = J_0 = J$ and zero field ($H = 0$). These restrictions will be released in the next section.

By symmetry arguments and without loss of generality, we restrict the rotation of all spins to the $yz$-plane, which means that the energy does not depend on the azimuthal angle. The orientation of the magnetic field is given by its polar angle $\theta_H$. Consequently, the Hamiltonian can now be expressed as

$$\begin{aligned}\frac{\mathcal{H}}{J} = &\, d_S \sum_{n=1}^{\mathcal{N}} \cos^2 \theta_n - d_H \cos^2 \theta_0 \\ &- \frac{1}{2} \sum_{n=1}^{\mathcal{N}-1} [\cos(\theta_n - \theta_{n+1}) + \cos(\theta_n - \theta_{n-1})] \\ &- \frac{1}{2} \cos(\theta_{\mathcal{N}} - \theta_{\mathcal{N}-1}) - \frac{1}{2} \cos(\theta_1 - \theta_0).\end{aligned} \quad (2)$$

where we have introduced the dimensionless parameters $d_H \equiv D_H/J$ and $d_S \equiv D_S/J$. The problem is thus mapped onto an effective spin chain with a pinned end at $n = 0$ where the spin is that of the HL, $i.e.$, $\mathbf{S}_0$, and a free at $n = \mathcal{N}$, $i.e.$, at the top of the SL.

In the following we present the analytical solution which turns out to be exact. Then the results for the magnetization profile are also compared to the numerical calculations.

#### A. Exact solution in the continuum

Minimizing the energy (2) with respect to the angle $\theta_n$ ($1 \leq n < \mathcal{N}$) taking account of the boundary conditions leads to the following equations

$$\begin{cases} \sin(\theta_n - \theta_{n+1}) + \sin(\theta_n - \theta_{n-1}) - d_S \sin(2\theta_n) = 0, \\ \text{for } 1 \leq n < \mathcal{N} \\ \sin(\theta_{\mathcal{N}} - \theta_{\mathcal{N}-1}) - d_S \sin 2\theta_{\mathcal{N}} = 0. \end{cases} \quad (3)$$

Next, for convenience, we introduce the parameter $\xi_n \equiv \theta_n - \theta_e$ where $\theta_e$ is the equilibrium polar angle of the atomic multilayer system. This is obtained by assuming that deep within the SL all layers spins are parallel to each other in the direction $\theta_e$, defined by the equation

$$\sin \theta_e - d_S \sin 2\theta_e + h \sin(\theta_e - \theta_H) = 0. \quad (4)$$

The first term stems from the exchange coupling to the HL. In the absence of the latter one obtains the usual Stoner-Wohlfarth equation of a macrospin in an oblique field. In fact, for typical materials $d_S \ll 1$ and thus $\theta_e \simeq 0$. Furthermore, assuming that the deviation between two consecutive spins is small and thereby expanding the equations in (3) with respect to $\xi_{n+1} - \xi_n \ll 1$, leads to the differential equations

$$\begin{cases} \frac{d^2\xi}{dz^2} + d_S \sin[2\xi(z)] = 0, \quad z \in [0, L[ \\ \left.\frac{d\xi}{dz}\right|_{z=L} - d_S \sin 2\xi_L = 0, \end{cases} \quad (5)$$

where $L$ is the thickness of the magnetic SL, $L = (\mathcal{N} - 1)a$. Here $z = 0$ corresponds to the first iron layer $n = 1$ and we can write $z_n = L \times \frac{n-1}{\mathcal{N}-1}$. $\xi_L$ is the value of the angle deviation at the free end at the position $z_L = L$.

We recognize in the first line of Eq. (5) the sine-Gordon equation with the Cauchy boundary condition in the second line. Solving it with the condition $\xi = 0$ at $z = 0$, we obtain

$$z = \int_0^\xi \frac{d\eta}{\sqrt{C_L + d_S \cos 2\eta}} = \frac{F(\xi, k)}{\sqrt{C_L + d_S}} \quad (6)$$

where $F(\xi_L, k)$ is the elliptic integral of the first kind whose module $k$ is given by

$$k^2 \equiv \frac{2d_S}{C_L + d_S}. \quad (7)$$

$C_L$ is the integration constant which depends on $\xi_L$ and is obtained from the first integral in Eq. (6) evaluated at $z = L$,

$$C_L = (d_S \sin 2\xi_L)^2 - d_S \cos 2\xi_L. \quad (8)$$

Inverting Eq. (6) yields the angle deviation $\xi$ as a function of the layer position $z$

$$\xi(z) = \arcsin\left[\mathbf{sn}\left(\sqrt{C_L + d_S} \times z\right)\right] \quad (9)$$

where $\mathbf{sn}$ is the Jacobi elliptic sine function.

This kind of profile was also obtained by Goto et al.[17] in the case of an extremely soft material on top of an extremely hard material. These authors dealt with the different issue of switching mechanisms in uniaxial films. An extension of this work to discrete multilayers with alternating hard and SL can be found in Ref. 18.

Next, the constant $\xi_L$ is obtained by integrating in Eq. (5) over $z$ from 0 to $L$ corresponding to $\xi$ ranging from 0 to $\xi_L$, thus leading to the equation

$$F\left(\arcsin[k \sin \xi_L], \frac{1}{k^2}\right) = L\sqrt{2d_S} \quad (10)$$

Note that a transformation has been done so that the module of the elliptic function becomes smaller than unity[19,20].

The overall problem is then solved in two steps: Eq. (8) is used to obtain $C_L$ in terms of $\xi_L$ and Eq. (10) yields $\xi_L$ as a function of the thickness $L$. Finally, inserting the results back into Eq. (9) yields the profile of the angle deviation $\xi$ as a function of the layer position $z$ for a given thickness $L$. This yields the exact solution of the problem (5)

Asymptotic expressions for the angle deviation $\xi(z)$ in Eq. (9) are then derived for layers near the SL/HL interface (small $\xi$) and for those approaching the surface layer (top of the SL), i.e. for supposedly large $\xi$.

For small $\xi$, an expansion with respect to the integrand $\eta$ in Eq. (6) yields

$$z = \frac{1}{\sqrt{C_L + d_S}} \int_0^\xi \frac{d\eta}{\sqrt{1 - (k\eta)^2}} \simeq \frac{\arcsin(k\xi)}{\sqrt{2d_S}}$$

which may be inverted and further expanded for small $z$. In addition, the use of Eq. (8) for $C(\xi_L)$ allows us to write explicitly

$$\xi(z) \simeq \left(\sqrt{2d_S} \sin \xi_L \sqrt{1 + 2d_S \cos^2 \xi_L}\right) \times z \quad (11)$$

This shows that the angle deviation is linear in the layer coordinate $z$ near the SL/HL interface.

For large $\xi$, with $\xi \lesssim \xi_L$, we write $\xi = \xi_L - \varepsilon$ with $\varepsilon$ being a small positive number. To first order in this expansion, we obtain the following asymptote

$$\xi(z) \simeq \xi_L - \sqrt{(1 - k^2 \sin^2 \xi_L)(C_L + d_S)} \times (L - z). \quad (12)$$

Going beyond this linear expression renders a cumbersome analysis without further relevant information about the physics of the problem.

The exchange coupling is at least two orders of magnitude larger than anisotropy, i.e., $d_S \sim 10^{-2}$, implying that within the exchange correlation length all spins $\mathbf{S}_n$ align along $\mathbf{S}_0$, itself tightly held by the hard anisotropy of the underlying material. However, as the length of the chain increases, a soft mode develops along the chain and induces spin deviations. Indeed, there is a minimal number of layers, $\mathcal{N}_{\min}$ necessary for the onset of noncolinearities of the spins $\mathbf{S}_n$. An estimation of $\mathcal{N}_{\min}$ can be obtained by comparing the exchange coupling energy to the in-plane anisotropy: assuming that the change in angle is uniform and achieved over $\mathcal{N}$ spins, the difference in exchange must equal, at $\mathcal{N} = \mathcal{N}_{\min}$, the anisotropy energy $|\Delta E_{\text{anis}}| = D_S S^2$, leading to

$$\mathcal{N}_{\min} \simeq \frac{\pi}{2\sqrt{2d_S}}. \quad (13)$$

For instance, for $d_S = 0.05$ we have $\mathcal{N}_{\min} \sim 5$ and for $d_S = 0.01$, $\mathcal{N}_{\min} \sim 11$.

For a typical anisotropy, $d_S = 0.01$, the main graph in Fig.2 displays the angle deviation $\xi(z)$ given by Eq. (9) for various values of $L$, the multilayer thickness. The corresponding asymptotes for small $z$ given by Eq. (11) and for large $z$ given by Eq. (12) are also plotted as dashed lines. The curve in circles is the angle deviation at the free boundary, as in the inset. Note that we use the same abscissa axis $z$ for both the curves $\xi(z)$ representing each a function of the layer coordinate $z \in [0, L]$ for a given multilayer thickness $L$, and the curve $\xi_L$ which is a function of $L$. The inset of Fig. 2 represents the value $\xi_L$ of the angle deviation at the loose boundary against the multilayer thickness $L$ as is given by Eq. (8), for $d_S = 0.01, 0.025, 0.05$. These curves cross the $z$ (or $L$) axis approximately at the number estimated by Eq. (13). It is seen that the stronger the in-plane anisotropy, the more rapidly $\xi_L$ converges to its largest value $\pi/2$ and the smaller the number of layers required for the onset of spin deviations.

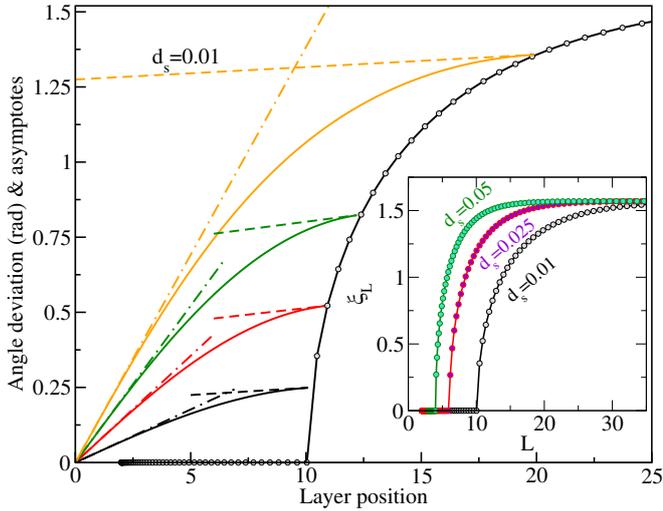

Figure 2: Angle deviation $\xi(z)$ as a function of the layer position $z$ within the SL for different values of the multilayer thickness $L$. The curve in circles is the angle deviation at the free boundary for $d_S = 0.01$. Inset: angle deviation at the free boundary against $L$, for $d_S = 0.01; 0.025; 0.05$.

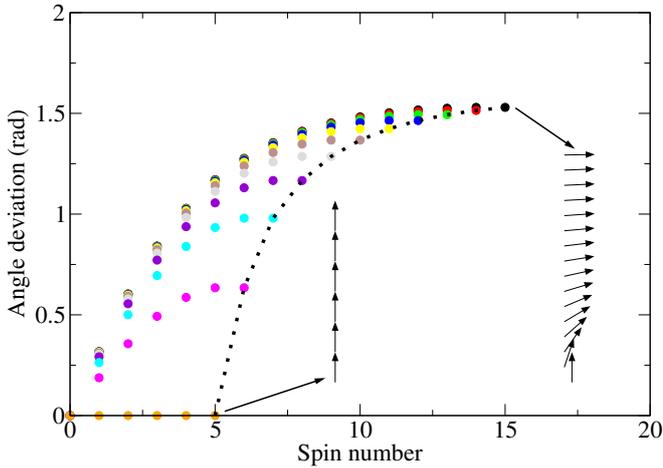

Figure 3: Angular deviation of the individual spins as a function of the number of spins in the Fe (soft magnetic) layer with $D_{Fe}/J_{Fe} = 0.05$. From left to right the two sets of arrows represent the rigid-magnet and ES regimes.

### B. Comparison with numerical calculations

Numerically, we simply minimize the energy (2) with respect to the angle $\theta_n$ taking account of the boundary conditions. This directly renders the angle $\theta_n$ as a function of the layer index $n$ or the layer position $z$ for a given multilayer thickness $L$ and a set of physical parameter, namely the exchange coupling, anisotropy, and applied field. In this case the problem is a discrete one, formed by the individual layer spins.

In Fig. 3 we plot the angle deviation as a function of spin position (number) for an anisotropy $d_S = 0.05$ as rendered by our numerical calculations. Here we clearly see the transition from the RM configuration to the ES regime at the $\mathcal{N}_{\min}$ given by Eq. (13). The spin configurations corresponding to the two regimes are indicated by arrows in the figure. We also note that the full $\pi/2$ deviation for the uppermost spin occurs for a system with more than 15 spins. Figure 4 represents a 2D false color map of these profiles.

In Fig. 5 we show a comparison between the analytical expression (9) and the results of the numerical simulations for $d_S = 0.05$. These plots show a perfect agreement between the (independent) analytical and numerical methods. It should be noted, however, that the analytical formula was obtained in the continuum limit as a function of the continuous layer position, whereas the numerical simulations use a discrete approach.

This favorable comparison provides a benchmark for the numerical method which is then applied to more realistic situations with the HL itself regarded as a multilayer system (relaxed interface) and in the presence of an oblique arbitrary magnetic field. These issues will be investigated in the following sections using the numerical simulation. The corresponding analytical developments together with the study of dynamical properties, will be studied in a subsequent work.

### IV. RELAXED INTERFACE

One of our objectives here is to understand how the previous system minimizes its energy as the now multi-layered HL absorbs the domain wall through the interface with the SL. Hence, the uppermost Fe layer can be seen as a well and the FePt hard multi-layer as a sink for the spins deviations. In particular, it is quite instructive to investigate the way the domain wall formation, its position and its width depend on the microscopic parameters and internal structure, such as the number of atomic planes in each film and the external applied field.

In the previous section, we considered the case of a HL as a massive ferromagnetic layer represented by a pinned macroscopic magnetic moment. Now we consider a more realistic situation, as shown in Fig.1, where both the Fe and FePt films have variable widths. In the analytical approach, as discussed in section II, one could label the whole set of atomic layers using the index $n = -\mathcal{M} \ldots \mathcal{N}$. In this case, the boundary condition at the HL/SL interface is replaced by continuity conditions.

In the following, we shall consider the situations with and without an applied magnetic field. The whole treatment here is done with the help of numerical simulations.

#### A. No magnetic field

In the absence of magnetic field, we study the effect of having a multilayered FePt film with the possibility for the corresponding spins to accommodate to the equilibrium conditions. In particular, it is instructive to figure out how the total energy of the system reduces as the soft mode that propagates through the SL comes down into the HL. Moreover, we note that in the situation with an equal number of spins on either side of the interface (*i.e.* inside SL and HL), the spin configuration is asymmetric; this is due to the differences in the exchange coupling constants of the SL and HL ($J_{Fe}$ and $J_{FePt}$) and their respective anisotropies. Indeed, the FePt film is more rigid and thereby the domain wall penetrates less into it than into the Fe film.

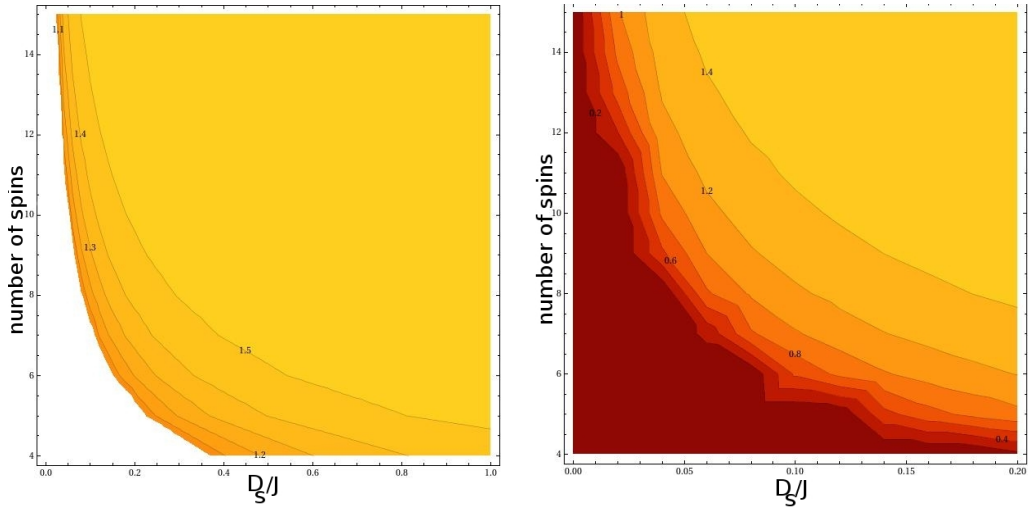

Figure 4: Uppermost spin representation of the simulation of the rigid magnet (RM) - exchange spring (ES) transition as a function of the Fe layer thickness (number of spins) and the $d_S = D_S/J$ ratio. The two figures show the same data with different scales.

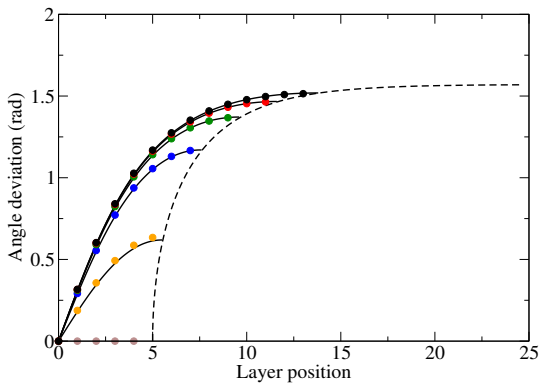

Figure 5: Full curves are the (analytical) angle deviation $\xi$ given in Eq. (9) as a function of the layer position $z$ for different values of the multilayer thickness $L$. Full circles are the numerical results. The dashed curve is the angle deviation at the free boundary against the multilayer.

From our simulations, we see that the profile of spin deviations across the bi-layer system is always smooth and continuous even across the interface. In Fig. 6 (Left) we show the analogous calculation for the RI conditions illustrated in Fig. 3. For the FePt layer with 6 spins and variable Fe layer thikness, we observe the RM-ES transition form 4-5 spins. Further increase of number of spins allows the system to relax and the DW is cleary seen to penetrate into the FePt layer. With the addition of spins in the Fe layer, the degree of accommodation of the domain wall increases and the spins further deviate from the perpendicular orientation ($\theta = 0$) in both hard and soft magnetic materials. As the number of Fe spins increases the spin orientation at the top layer gradually (asymptotically) reaches the in-plane orientation ($\pi/2$). This is clearly seen in Fig. 6. Once again a transition RM/ES is observed, though it is more gradual than in the case of a rigid interface and occurs for less Fe atomic planes. Furthermore, we note that even for a very few planes in the Fe layer a small relaxation is observed.

By increasing the thickness of the FePt film, the domain wall is allowed to relax further into the HL, as shown in Fig. 7. The gradual rotation of the ES domain wall is much slower in this case since the exchange energy is distributed along a longer spin chain. Therefore the full 90° rotation will be asymptotically reached for even thicker Fe films.

Varying the anisotropy constant $D_H$ in the HL also has an important influence on the spin configuration: as the value of $D_H$ increases the FePt spins become more closely aligned along the perpendicular direction. In this context the rigid interface regime corresponds to the case of infinite $D_H$. It is clear that the profile of the spin deviations across the interface undergoes changes in gradient which increase with the ratio $D_H/D_S$.

As a comparison with our calculations, we have used the OOMMF software package to perform analogous simulations. We have used the method of conjugate gradient with no pre-condition for simulating a sample with a size of $(1 \times 25 \times 10^{-4} \times 1)\mu$m and $(0.1 \times 10^{-5} \times 0.1)\mu$m. The exchange coupling between Fe/FePt (at the interface) was taken about 1.44 times larger than the exchange cou-pling within the Fe film. The main features of our model are reproduced with some minor differences. From the graphs illustrated in Fig. 8, we see that the same general features are reproduced with similar spin profiles.

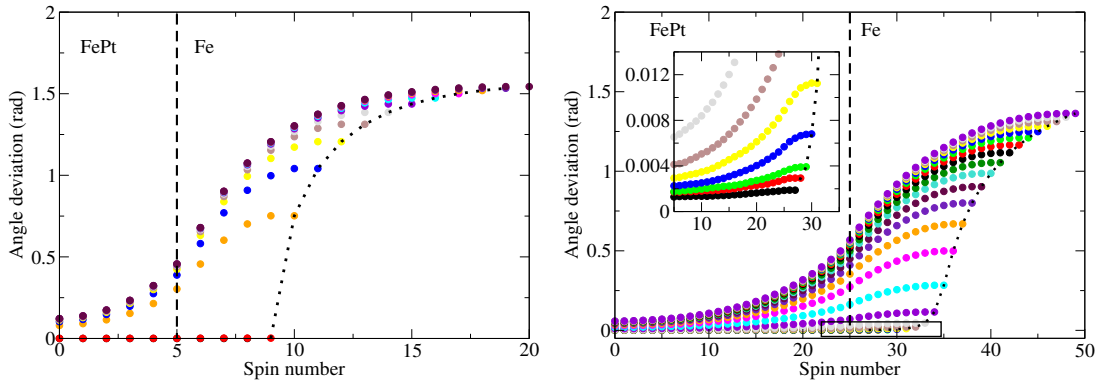

Figure 6: Angle deviation of the individual spins as a function of the number of spins in the Fe layer (relaxed interface) with a fixed number of spins in the FePt layer, for $J_{\text{FePt}}/J_{\text{Fe}} = 1.44$. (Left) FePt layer with 6 spins and $D_{\text{Fe}}/J_{\text{Fe}} = 0.1$, $D_{\text{FePt}}/J_{\text{Fe}} = 0.2$. We can see the transition from the RM regime to the ES regime between 4 and 5 atomic layers of Fe. (Right) FePt layer with 25 spins and $D_{\text{Fe}}/J_{\text{Fe}} = 0.01$, $D_{\text{FePt}}/J_{\text{Fe}} = 0.02$. We can see that the transition from the RM to the ES regime is more gradual than the one on the left. The inset shows an expanded view of the region indicated in the main graph by a rectangle.

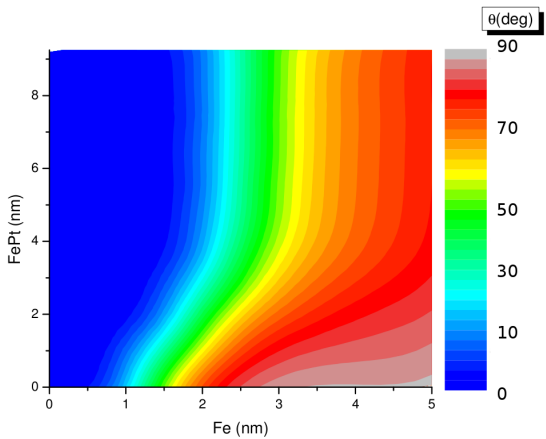

Figure 7: False color plot of the angle variation as a function of the thickness of the Fe film and the thickness of the FePt film.

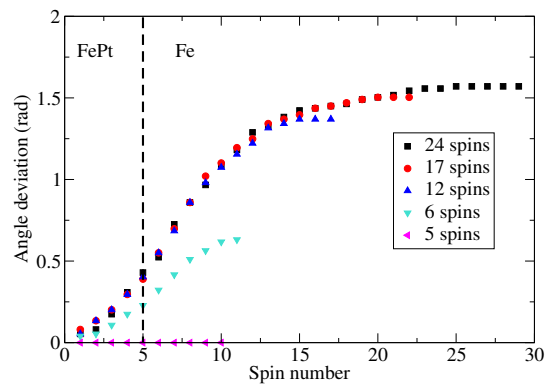

Figure 8: OOMMF simulations in the case of a relaxed interface; the FePt layer has 5 spins, while that for the Fe layer is varied.

### B. Effects of a magnetic field

We now consider the effect of an applied external magnetic field (in strength and direction) on the equilibrium spin configurations of our FePt/Fe bi-layer system. The results of our simulations for 25 atomic layers of FePt and 20 atomic layers of Fe are summarized in Fig. 9.

In the simplest case, we study the influence of a magnetic field on the equilibrium configuration when it is applied in the normal direction [see Fig. 9(a)]. In this case the applied field is parallel to the easy axis of the HL and has the effect of aligning all spins in the perpendicular direction. This can also be thought of as effectively pushing the domain wall up and eventually out of the layers.

The case of a field applied in the $45^o$ direction is rather peculiar (Fig. 9(b)) in a sense that as the field increases, the noncolinearities of spins reduce, $i.e.$ the domain wall is effectively compressed on both sides. This is to be expected since, upon increasing the field strength, those spins with an equilibrium orientation less than $45^o$ (in zero field) will increase their angle while those with an orientation greater than $45^o$ will reduce theirs. The fixed point clearly corresponds to $45^o$ and is unaffected by the change of the field magnitude. This amounts to a gradual alignment of the spins along the applied field direction. In fact this will be the general situation for the other orientations as well, where the spin distribution across the sample can be stretched or compressed, depending on the field direction and magnitude.

As the field turns into the film plane, the domain wall shifts further into the FePt layer, especially for stronger fields. This trend continues as the field is rotated further and below the film plane. In Fig. 9(e), the field is in the 180° direction, opposite to the easy axis of the FePt layer, there appears a slowing down of this trend, which can be expected since there is no in-plane component of the applied magnetic field.

We have also performed a series of simulations for different thicknesses of Fe and FePt layers with the same external conditions mentioned above. From these simulations we can infer a few instructive general results: we can see extremes for the overall behavior depending on which of the layers, HL or SL, is thicker. The results are intuitively easy to understand. For example, for the case with a thicker FePt film, the sample is dominated by the HL, as we have seen in the case of a sample with 25 FePt spins and 10 Fe spins. In the opposite situation with a sample of 5 FePt and 20 Fe planes, the properties of the SL dominate and a much stronger in-plane component is observed.

## V. EXPERIMENTAL CONTEXT AND PERSPECTIVES

The present work stems from previous experimental studies of the FePt/Fe bilayer system using the ferromagnetic resonance (FMR) technique[13]. In this work we were able to observe the effects of exchange coupling *via* an angular study of the resonance condition. Additionally, we observed an unexpected resonance feature, which appeared to have an off-perpendicular easy-axis. The spectra for rigid magnet and exchange spring regimes differed in the number of resonances obtained. In order to understand these dynamic properties it is necessary to understand the equilibrium configuration from which the dynamic situation evolves, and the theoretical work in this paper goes a long way to achieving this purpose. In a future publication we will develop the theory of the excitation modes (frequency – field characteristics) of this system, which should pave the way for a fuller interpretation of experimental work. In Fig. 7 we show the spin orientation for the FePt/Fe bilayer system as a function of the thicknesses of the two layers. Considering previous experimental studies of this system[10,21], it is instructive to compare the values obtained for the critical thickness of the soft (Fe) layer for which the transition from rigid magnet to exchange spring occurs. Indeed from the figure it will be noted that the most sensitive region occurs for an FePt thickness of less than 4 nm, above which the transition appears to be insensitive to the HL thickness. In the study of Casoli *et al.*[21] where the FePt layer of 10 nm is deposited on MgO (100) substrates, a transition to the ES state occurs above 2 nm of Fe and a fully ES configuration is obtained for 3.5 nm. Since the transition thickness is influenced by the interface morphology and extrinsic properties an accurate value can only be obtained by a theoretical study. However, from Fig. 7 we see that for this region of FePt thickness (*i.e.* the upper limit of our figure) that we expect the transition to initiate at around 2 nm of Fe. Furthermore, our calculations show that for a thickness of 3.5 nm, the Fe upper spins have an orientation of around 60°. This is more than sufficient to produce the exchange spring like behavior observed experimentally. It is worth stressing that the resolution of Fig. 7 is such that very small deviations from the perpendicular will not be observed, *cf.* inset of Fig. 6(Right), where for very small Fe thicknesses small deviations from the perpendicular are evident. A more sensitive experimental measurement of the spin profile in bilayer systems is required to make a more reliable comparison between experiment and theory. One possible method would be to produce a sample in which both the HL and SL are wedge shaped with a cross wedge structure. Mapping the Fe surface magnetisation orientation should then enable the direct visualization of the RM – ES transition as a function of both layer thicknesses.

In future work we aim to study the dynamic properties of the FePt/Fe bilayer structure both from the theoretical as well as the experimental point of view. We intend to extend our model for the energy using the Landau – Lifshitz formalism to obtain the spin dynamics on a spin by spin basis and also take into account the boundary conditions. Indeed we have already performed dynamic simulations on this system for the rigid interface conditions using the OOMMF software, which provides some insights [22].

In this paper we predict the existence of multipeaked (frequency) spectra, which for perpendicular applied magnetic fields are of a regular nature. We have excluded the possibility of spin wave excitations and believe that these arise from the varying effective internal field across the thickness of the sample, representing the different equilibrium conditions in the ES state. Experimentally, measurements could be made using ferromagnetic resonance (FMR) or with the aid of a network analyzer (NA). Indeed, NA-FMR could provide an excellent tool for a comparison of the frequency spectra we calculate since frequency sweeps at fixed applied fields can be directly measured.

## VI. CONCLUSION AND OUTLOOK

We have made an extensive theoretical study of the HL/SL bilayer coupled system, where we have used FePt (HL) and Fe (SL) as a model system since we have previously studied this system experimentally. We have considered the effect of the interface pinning by considering the cases of a rigid and a relaxed interface between HL and SL. In the former case analytical and numerical calculations have shown that the magnetisation profile is a smooth function and can be expressed by the Jacobi elliptic sine function. The excellent agreement between the analytical approach in the continuum limit and the numerical simulations performed for a discrete lattice suggests that for the system studied the continuum is reached with a relatively small number of layers.

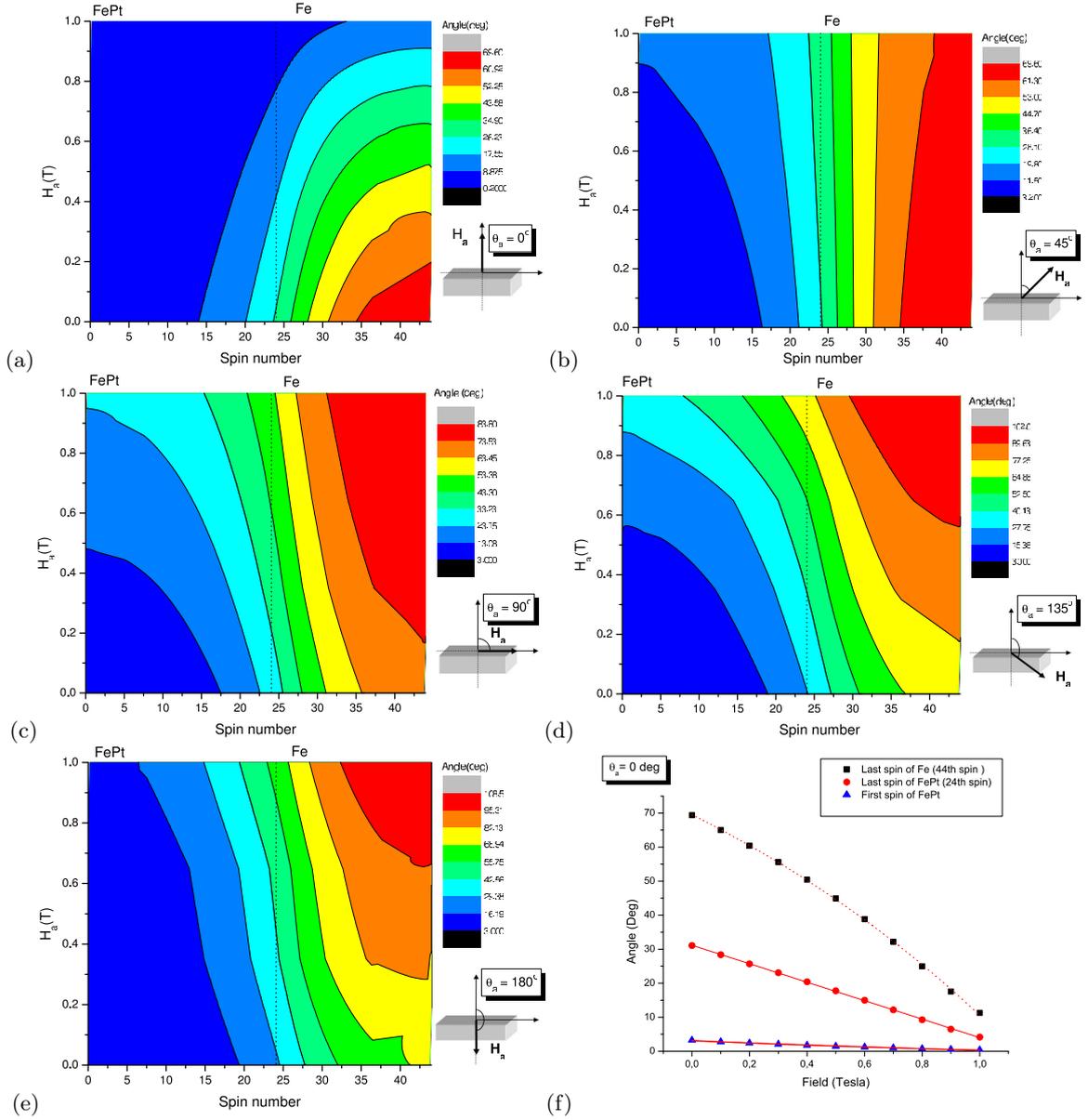

Figure 9: Variation of the spin configuration as a function of applied magnetic field. (a) - (e) show the effect of field direction. Note that (a) corresponds to the same data as in Fig. 6(Right). (f) Spin angle as a function of applied magnetic field for spins at the FePt outer surface (blue triangles), at the Fe - FePt interface (red circles) and at the outer Fe surface (black squares).

The rigid interface condition essentially treats the HL as a macrospin with fixed perpendicular orientation. Both analytical and numerical calculations predict a RM – ES transition for the FePt/Fe system from between 5 and 6 atomic (spins) layers, which corresponds to a layer thickness of around 1 nm. In the case of the relaxed interface, we have used numerical simulations to obtain the stable equilibrium conditions. This shows a more gradual RM – ES transition: this can be seen from Fig. 6, where the transition is seen to be broader for thicker HL. From 4 nm, there seems to be very little change in the transition. OOMMF simulations give very similar results to those obtained using our model. The overall magnetic properties of the bilayer structure will depend on the thicknesses of the two individual layers and will be dominated by the thicker layer in general. We have further considered the effects of an external magnetic field on the equilibrium configuration, which we have applied as a function of field strength and direction. In general, the magnetic field can compress, expand and even eliminate the domain wall structure from the sample. Our basic model is a very general one, and should be applicable to any bilayer system for which the exchange and anisotropy constants are known. The comparison of our theoretical results seems to be favourable with experimental results in the FePt/Fe system studied.

**Acknowledgments**

We are grateful for bilateral exchange grants awarded between the University of Porto and IMEM (FCT, Portugal - CNR, Italy) and the University of Porto and the University of Perpignan (FCT, Portugal - Egide, CNRS, France). A. Apolinario acknowledges FCUP (Faculty of Sciences of Oporto University) the financial support under the FCT project PTDC/FIS/66262/2006. H.K. acknowledges a useful discussion of elliptic functions with P. M. Déjardin.

---